\documentclass[twocolumn,showpacs,preprintnumbers,superscriptaddress]{revtex4}%
\usepackage{amssymb}
\usepackage{amsmath}
\usepackage{graphicx}
\usepackage{dcolumn}
\usepackage{bm}
\usepackage{amsfonts}
\usepackage{epstopdf}%
\providecommand{\U}[1]{\protect\rule{.1in}{.1in}}
\providecommand{\U}[1]{\protect\rule{.1in}{.1in}}
\begin{document}
\title{Exploring the nonlinear regime of light-matter interaction using electronic spins in diamond}
\author{Nir Alfasi*\footnotetext{*These authors contributed equally to this work.}}
\affiliation{Andrew and Erna Viterbi Department of Electrical Engineering, Technion, Haifa
32000 Israel}
\author{Sergei Masis*}
\affiliation{Andrew and Erna Viterbi Department of Electrical Engineering, Technion, Haifa
32000 Israel}
\author{Roni Winik}
\affiliation{Andrew and Erna Viterbi Department of Electrical Engineering, Technion, Haifa
32000 Israel}
\author{Demitry Farfurnik}
\affiliation{Racah Institute of Physics, The Hebrew University of Jerusalem, Jerusalem
9190401, Israel}
\author{Oleg Shtempluck}
\affiliation{Andrew and Erna Viterbi Department of Electrical Engineering, Technion, Haifa
32000 Israel}
\author{Nir Bar-Gill}
\affiliation{Racah Institute of Physics, The Hebrew University of Jerusalem, Jerusalem
9190401, Israel}
\author{Eyal Buks}
\affiliation{Andrew and Erna Viterbi Department of Electrical Engineering, Technion, Haifa
32000 Israel}
\date{\today}

\begin{abstract}
The coupling between defects in diamond and a superconducting microwave
resonator is studied in the nonlinear regime. Both negatively charged
nitrogen-vacancy and P1 defects are explored. The measured cavity mode
response exhibits strong nonlinearity near a spin resonance. Data is compared
with theoretical predictions and a good agreement is obtained in a wide range of externally controlled parameters. The nonlinear
effect under study in the current paper is expected to play a role in any
cavity-based magnetic resonance imaging technique and to impose a fundamental
limit upon its sensitivity.

\end{abstract}
\pacs{42.50.Pq,81.05.ug,76.30.Mi}
\maketitle

%Force line breaks with \\

%Lines break automatically or can be forced with \\

%It is always \today, today,
%but any date may be explicitly specified

%PACS, the Physics and Astronomy
%Classification Scheme.
%\keywords{Suggested keywords}%Use showkeys class option if keyword
%display desired

Cavity quantum electrodynamics (CQED) \cite{Haroche_24} is the study of the
interaction between photons confined in a cavity and matter. CQED has
applications in a variety of fields, including magnetic resonance imaging and
quantum computation \cite{Wallraff2004}. The CQED interaction can be probed by
measuring the response of a cavity mode. Commonly, the effect of matter on the
response diminishes as the energy stored in the cavity mode under study is
increased \cite{Anders_NonlinearESR}. This nonlinear effect, which is the
focus of the current study, imposes a severe limit upon the performance of a
variety of CQED systems.

In the current study we explore nonlinear CQED interaction between defects in
a diamond crystal and a superconducting microwave cavity (resonator) having a
spiral shape \cite{Maleeva_474,Maleeva_064910}. Two types of defects are
investigated, a negatively charged nitrogen-vacancy NV$^{-}$ defect and a
nitrogen 14 (nuclear spin 1) substitutional defect (P1). Strong coupling
between defects in diamond and a superconducting resonator has been
demonstrated at ultra-low temperatures
\cite{Zhu2011,Kubo_140502,Kubo_220501,Amsuss_060502,Schuster_140501,Sandner_053806,Grezes_021049}%
, however the regime of nonlinear response was not addressed. In this study,
we find that the cavity response becomes highly nonlinear near a CQED
resonance. In addition, for the case of NV$^{-}$ defects, the response is
strongly affected by applying optically-induced spin polarization (OISP). The
experimental findings are compared with theory and good agreement is obtained.%

%TCIMACRO{\FRAME{ftbpFU}{3.4537in}{1.7877in}{0pt}{\Qcb{The experimental setup.
%(a) A loop antenna (LA) is coupled to the spiral resonator. Two multimode
%optical fibers are coupled to the diamond wafer. Fiber F1 is employed for
%delivering laser light of wavelength $\lambda_{\mathrm{L}}=532\unit{nm}$, and
%fiber F2 probes the emitted photoluminescence (PL). (b) The spiral resonator
%has $3$ turns, an inner radius of $0.59\unit{mm}$ and an outer radius of
%$0.79\unit{mm}$. (c) The resonance lineshape of the spiral's fundamental mode
%vs. temperature. (d) The magnetic induction magnitude $\left\vert
%\mathbf{B}_{\mathrm{c}}\left(  \mathbf{r}\right)  \right\vert $ of the
%fundamental mode vs. position $\mathbf{r}$ in a plane perpendicular to both
%wafers that contains the center of the spiral.}}{\Qlb{Fig_resonator}%
%}{figresonator.eps}{\special{ language "Scientific Word";  type "GRAPHIC";
%maintain-aspect-ratio TRUE;  display "ICON";  valid_file "F";
%width 3.4537in;  height 1.7877in;  depth 0pt;  original-width 25.1963in;
%original-height 12.9413in;  cropleft "0";  croptop "1";  cropright "1";
%cropbottom "0";  filename 'FigResonator.eps';file-properties "XNPEU";}} }%
%BeginExpansion
\begin{figure}
[ptb]
\begin{center}
\includegraphics[
height=1.7877in,
width=3.4537in
]%
{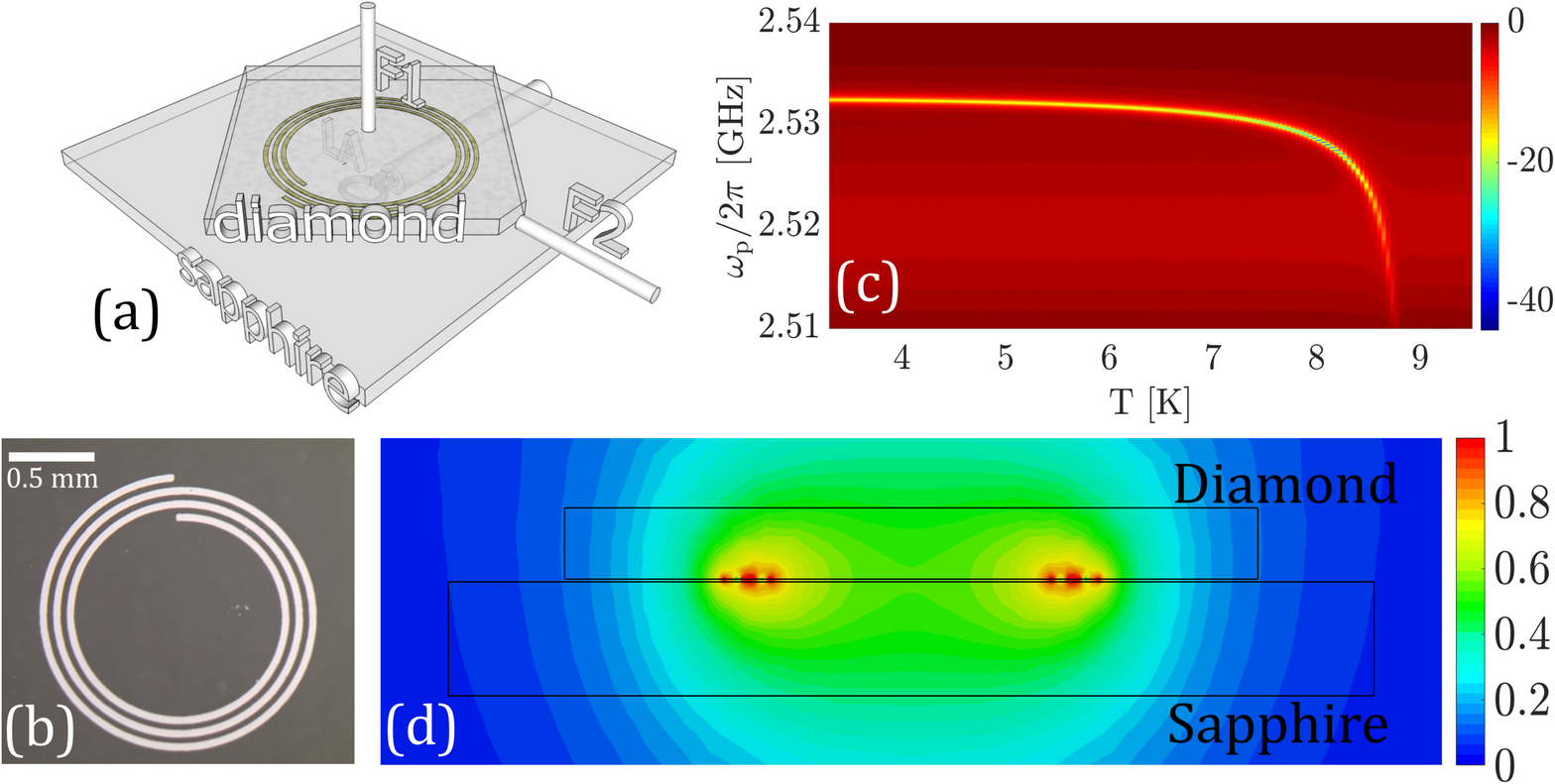}%
\caption{The experimental setup. (a) A loop antenna (LA) is coupled to the
spiral resonator. Two multimode optical fibers are coupled to the diamond
wafer. Fiber F1 is employed for delivering laser light of wavelength
$\lambda_{\mathrm{L}}=532\operatorname{nm}$, and fiber F2 probes the emitted
photoluminescence (PL). (b) The spiral resonator has $3$ turns, an inner
radius of $0.59\operatorname{mm}$ and an outer radius of
$0.79\operatorname{mm}$. (c) The resonance lineshape of the spiral's
fundamental mode vs. temperature. (d) The magnetic induction magnitude
$\left\vert \mathbf{B}_{\mathrm{c}}\left(  \mathbf{r}\right)  \right\vert $ of
the fundamental mode vs. position $\mathbf{r}$ in a plane perpendicular to
both wafers that contains the center of the spiral.}%
\label{Fig_resonator}%
\end{center}
\end{figure}
%EndExpansion

\begin{figure}
[ptb]
\begin{center}
\includegraphics[
height=1.904in,
width=3.4546in
]%
{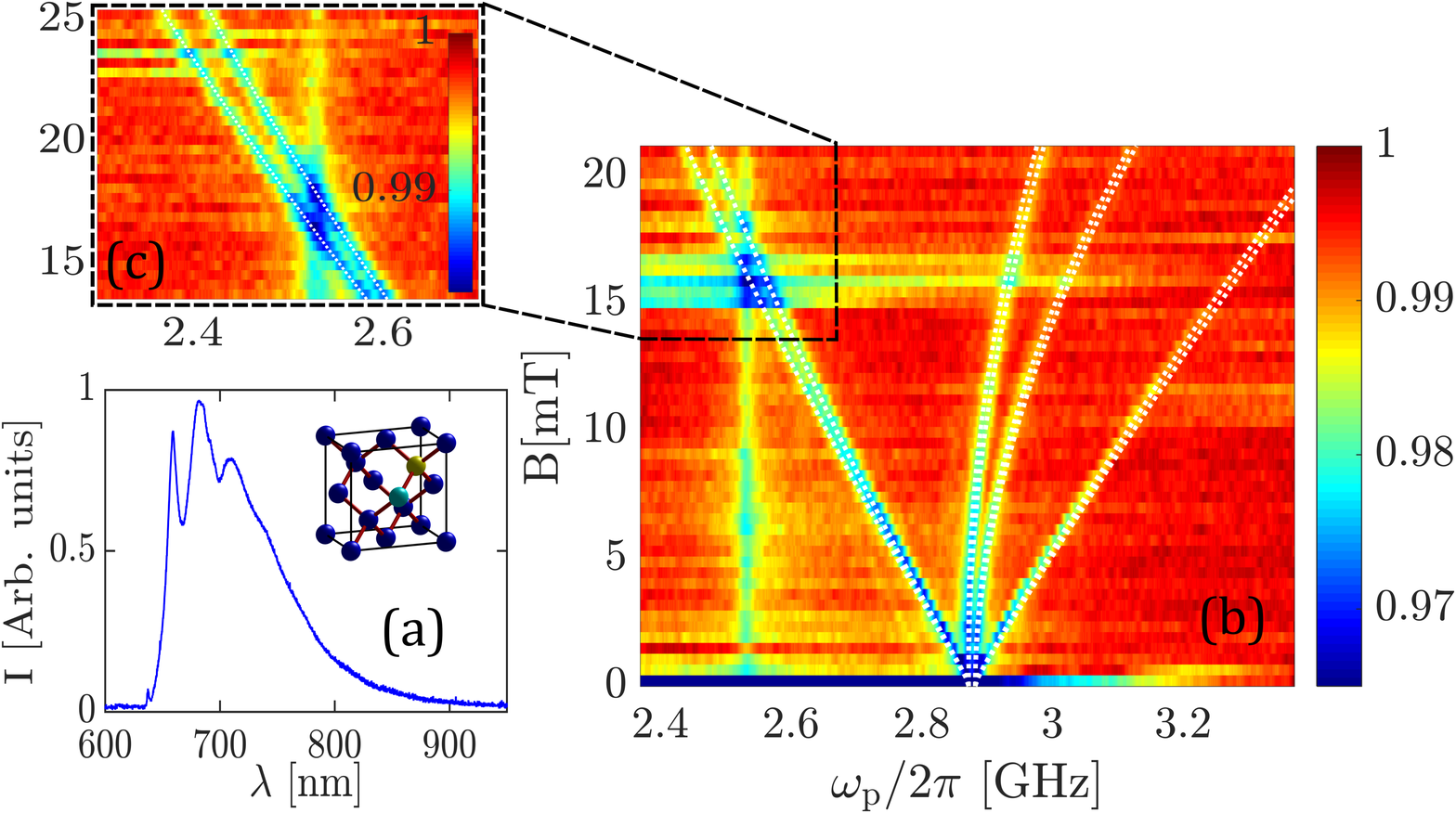}%
\caption{ODMR (a) The measured emitted PL spectrum. (b) and (c) ODMR spectrum
vs. driving frequency $\omega_{\mathrm{p}}/2\pi$ and magnetic field
$\left\vert \mathbf{B}\right\vert $. The white dotted lines represent the
frequencies $\omega_{\pm}/2\pi$ calculated using Eq.~(\ref{omega pm}). The
fitting procedure yields the direction of the magnetic field $\mathbf{\hat{b}%
}$, which is expressed as $\mathbf{\hat{b}}=T_{\mathbf{\hat{z}}}\left(
\theta_{z}\right)  T_{\mathbf{\hat{y}}}\left(  \theta_{y}\right)
T_{\mathbf{\hat{x}}}\left(  \theta_{x}\right)  \mathbf{\hat{z}}$, where
$\mathbf{\hat{x}}$, $\mathbf{\hat{y}}$ and $\mathbf{\hat{z}}$ are unit vectors
in the crystal directions $\left[  100\right]  $, $\left[  010\right]  $ and
$\left[  001\right]  $, respectively, and $T_{\mathbf{\hat{s}}}\left(
\theta_{s}\right)  $ is a rotation operator around the axis $\mathbf{\hat{s}}%
$, where $s\in\left\{  x,y,z\right\}  $. The rotation angles found from the
fitting procedure are $\theta_{x}/\pi=-0.02$, $\theta_{y}/\pi=0.002$ and
$\theta_{z}/\pi=0.05$.}%
\label{Fig_diamond}%
\end{center}
\end{figure}

The experimental setup is schematically depicted in Fig.~\ref{Fig_resonator}%
(a). Defects in a [100] type Ib diamond are created using $2.8%
%TCIMACRO{\unit{MeV}}%
%BeginExpansion
\operatorname{MeV}%
%EndExpansion
$ electron irradiation with a dose of approximately $8\times10^{18}%
~\mathrm{e/cm^{2}}$, followed by annealing at $800^{\circ}\mathrm{C}$ for 8
hours and acid cleaning, resulting in the formation of NV$^{-}$ defects with
density of $1.23\times10^{17}%
%TCIMACRO{\unit{cm}}%
%BeginExpansion
\operatorname{cm}%
%EndExpansion
^{-3}$ \cite{Farfurnik_123101}. The diamond wafer is then placed on top of a
sapphire wafer supporting a superconducting spiral resonator made of niobium
[see Fig.~\ref{Fig_resonator}(b)]. Externally applied magnetic field
$\mathbf{B}$ is employed for tuning the system into a CQED resonance. A
coaxial cable terminated by a loop antenna (LA) transmits both injected and
off-reflected microwave signals. The LA has a coupling given by $\gamma
_{\mathrm{f}}/2\pi=0.367%
%TCIMACRO{\unit{MHz}}%
%BeginExpansion
\operatorname{MHz}%
%EndExpansion
$ to the spiral's fundamental mode, which has a frequency of $\omega
_{\mathrm{c}}/2\pi=2.53%
%TCIMACRO{\unit{GHz}}%
%BeginExpansion
\operatorname{GHz}%
%EndExpansion
$ and an unloaded damping rate of $\gamma_{\mathrm{c}}/2\pi=0.253%
%TCIMACRO{\unit{MHz}}%
%BeginExpansion
\operatorname{MHz}%
%EndExpansion
$ [these values are extracted from a fitting based on Eq.~(\ref{R SNL})
below]. All measurements are performed at a base temperature of $T=3.1%
%TCIMACRO{\unit{K}}%
%BeginExpansion
\operatorname{K}%
%EndExpansion
$. A network analyzer (NA) measurement of the temperature dependence of the
resonance lineshape is seen in Fig.~\ref{Fig_resonator}(c). The color-coded
plot depicts the reflectivity coefficient $R_{\mathrm{c}}=P_{\mathrm{r}%
}/P_{\mathrm{p}}$ is dB units, where $P_{\mathrm{p}}=-70$ dBm and
$P_{\mathrm{r}}$ are, respectively, the injected power into the LA and the
off-reflected power from the LA, as a function of both frequency of injected
signal $\omega_{\mathrm{p}}/2\pi$ and temperature $T$. Laser light of
wavelength $\lambda_{\mathrm{L}}=532%
%TCIMACRO{\unit{nm}}%
%BeginExpansion
\operatorname{nm}%
%EndExpansion
$\ and intensity $I_{\mathrm{L}}$ (in units of power per unit area) is
injected into the diamond wafer using a multimode optical fiber F1, and
another multimode optical fiber F2 delivers the emitted photoluminescence (PL)
to an optical spectrum analyzer [see Fig.~\ref{Fig_diamond}(a)]. Numerical
calculation is employed for evaluating the shape of the spiral's fundamental
mode [see Fig.~\ref{Fig_resonator}(d)].

The negatively-charged $\text{NV}^{-}$ defect in diamond consists of a
substitutional nitrogen atom (N) combined with a neighbor vacancy (V)
\cite{Doherty_1}. The ground state of the NV$^{-}$ defect is a spin triplet
having symmetry $^{3}A_{2}$ \cite{maze_025025,wrachtrup2006}, composed of a
singlet state $\left\vert m_{\mathrm{e}}=0\right\rangle $ and a doublet
$\left\vert m_{\mathrm{e}}=\pm1\right\rangle $. The angular resonance
frequencies $\omega_{\pm}$ corresponding to the transitions between the state
$\left\vert m_{\mathrm{e}}=0\right\rangle $ and the states $\left\vert
m_{\mathrm{e}}=\pm1\right\rangle $ are approximately given by
\cite{Ovartchaiyapong_1403_4173,MacQuarrie_227602,Rondin_056503}%
\begin{equation}
\omega_{\pm}=D\pm\sqrt{\gamma_{\mathrm{e}}^{2}B_{\parallel}^{2}+E^{2}}%
+\frac{3}{2}\frac{\gamma_{\mathrm{e}}^{2}B_{\perp}^{2}}{D}\;, \label{omega pm}%
\end{equation}
where $B_{\parallel}$ is the magnetic field component parallel to the axis of
the NV defect and $B_{\perp}$ is the transverse one. The parameter
$\gamma_{\mathrm{e}}=2\pi\times28.03%
%TCIMACRO{\unit{GHz}}%
%BeginExpansion
\operatorname{GHz}%
%EndExpansion%
%TCIMACRO{\unit{T}}%
%BeginExpansion
\operatorname{T}%
%EndExpansion
^{-1}$ is the electron spin gyromagnetic ratio. In the absence of strain and
when the externally applied magnetic field vanishes one has $\omega_{\pm}=D$,
where $D=2\pi\times2.87%
%TCIMACRO{\unit{GHz}}%
%BeginExpansion
\operatorname{GHz}%
%EndExpansion
$. Internal strain, however, may lift the degeneracy between the states
$\left\vert m_{\mathrm{e}}=-1\right\rangle $ and $\left\vert m_{\mathrm{e}%
}=+1\right\rangle $, and give rise to a splitting given by $2E$ (in our sample
$E=2\pi\times10%
%TCIMACRO{\unit{MHz}}%
%BeginExpansion
\operatorname{MHz}%
%EndExpansion
$). In a single crystal diamond the NV defects have four different possible
orientations with four corresponding pairs of angular resonance frequencies
$\omega_{\pm}$.

The technique of optical detection of magnetic resonance (ODMR) can be
employed for measuring the resonance frequencies $\omega_{\pm}$
\cite{Gruber_2012,le_121202}. The measured PL spectrum is seen in
Fig.~\ref{Fig_diamond}(a). The integrated PL signal in the band $660%
%TCIMACRO{\unit{nm}}%
%BeginExpansion
\operatorname{nm}%
%EndExpansion
-760%
%TCIMACRO{\unit{nm}}%
%BeginExpansion
\operatorname{nm}%
%EndExpansion
$ is plotted as a function of microwave input frequency $\omega_{\mathrm{p}%
}/2\pi$ and externally applied magnetic field $\left\vert \mathbf{B}%
\right\vert $ in Figs.~\ref{Fig_diamond}(b)-(c). In this measurement the
microwave input power is set to $P_{\mathrm{p}}=20\;$dBm. The direction of the
externally applied magnetic field $\mathbf{B}$ is found by fitting the
measured ODMR frequencies $\omega_{\pm}$ with the calculated values given by
Eq.~(\ref{omega pm}).

The ODMR spectrum contains a profound resonance feature at the frequency of
the spiral resonator $\omega_{\mathrm{c}}/2\pi=2.53%
%TCIMACRO{\unit{GHz}}%
%BeginExpansion
\operatorname{GHz}%
%EndExpansion
$ [see Fig.~\ref{Fig_diamond}(b)]. This feature is attributed to
heating-induced change in the internal stress in the diamond wafer. Two (out
of four) resonance frequencies $\omega_{-}/2\pi$ can be tuned close to the
spiral resonator frequency $\omega_{\mathrm{c}}/2\pi$ by setting the magnetic
field $\left\vert \mathbf{B}\right\vert $ close to the value of $16%
%TCIMACRO{\unit{mT}}%
%BeginExpansion
\operatorname{mT}%
%EndExpansion
$. The two groups of NV$^{-}$ defects giving rise to these two resonances have
the smallest angles with respect to the externally applied magnetic field (see
caption of Fig. \ref{Fig_diamond}). In this region, which is magnified in
Fig.~\ref{Fig_diamond}(c), the deepest ODMR is obtained when the magnetic and
resonator frequencies coincide.%

%TCIMACRO{\FRAME{ftbpFU}{3.4546in}{1.904in}{0pt}{\Qcb{ODMR (a) The measured
%emitted PL spectrum. (b) and (c) ODMR spectrum vs. driving frequency
%$\omega_{\mathrm{p}}/2\pi$ and magnetic field $\left\vert \mathbf{B}%
%\right\vert $. The white dotted lines represent the frequencies $\omega_{\pm
%}/2\pi$ calculated using Eq.~(\ref{omega pm}). The fitting procedure yields
%the direction of the magnetic field $\mathbf{\hat{b}}$, which is expressed as
%$\mathbf{\hat{b}}=T_{\mathbf{\hat{z}}}\left(  \theta_{z}\right)
%T_{\mathbf{\hat{y}}}\left(  \theta_{y}\right)  T_{\mathbf{\hat{x}}}\left(
%\theta_{x}\right)  \mathbf{\hat{z}}$, where $\mathbf{\hat{x}}$, $\mathbf{\hat
%{y}}$ and $\mathbf{\hat{z}}$ are unit vectors in the crystal directions
%$\left[  100\right]  $, $\left[  010\right]  $ and $\left[  001\right]  $,
%respectively, and $T_{\mathbf{\hat{s}}}\left(  \theta_{s}\right)  $ is a
%rotation operator around the axis $\mathbf{\hat{s}}$, where $s\in\left\{
%x,y,z\right\}  $. The rotation angles found from the fitting procedure are
%$\theta_{x}/\pi=-0.02$, $\theta_{y}/\pi=0.002$ and $\theta_{z}/\pi=0.05$.}%
%}{\Qlb{Fig_diamond}}{figdiamond.eps}{\special{ language "Scientific Word";
%type "GRAPHIC";  maintain-aspect-ratio TRUE;  display "ICON";
%valid_file "F";  width 3.4546in;  height 1.904in;  depth 0pt;
%original-width 24.319in;  original-height 13.313in;  cropleft "0";
%croptop "1";  cropright "1";  cropbottom "0";
%filename 'FigDiamond.eps';file-properties "XNPEU";}} }%
%BeginExpansion
%EndExpansion

\begin{figure*}
[ptb]
\begin{center}
\includegraphics[
height=5.5052in,
width=6.9074in
]%
{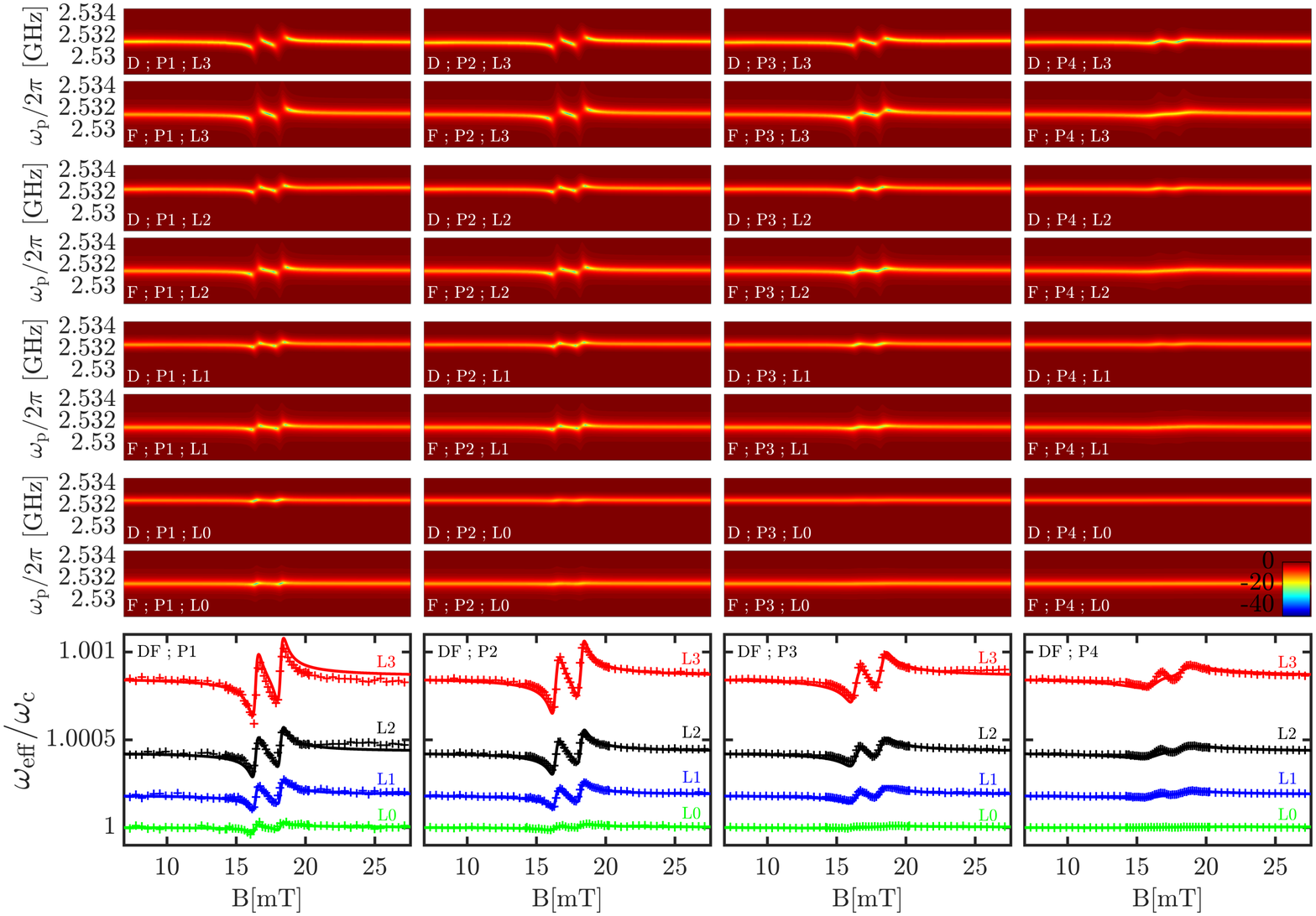}%
\caption{Cavity mode reflectivity $R_{\mathrm{c}}$ with NV$^{-}$ defects for
various values of injected microwave power $P_{\mathrm{p}}$ ($\mathrm{P}%
1=-90~\mathrm{dBm}$, $\mathrm{P}2=-70~\mathrm{dBm}$, $\mathrm{P}%
3=-60~\mathrm{dBm}$ and $\mathrm{P}%
4=-50~\mathrm{dBm}$) and laser power $I_{\mathrm{L}}$ ($\mathrm{L}0=0$,
$\mathrm{L}1=5.6\operatorname{mW}\operatorname{mm}^{-2}$, $\mathrm{L}%
2=12.8\operatorname{mW}\operatorname{mm}^{-2}$ and $\mathrm{L}%
3=30\operatorname{mW}\operatorname{mm}^{-2}$). For each pair the top plot is
experimental data (labeled by D) and the bottom is the theoretical prediction
(labeled by F). The bottom row shows the effective normalized resonance frequency ($\omega_\text{eff}/\omega_\text{c}$), which is obtained from the minimum reflectivity signal for each magnetic field, as a function of magnetic field. Each column corresponds to a single microwave power (DF ; P1-P4) and each plot shows all four laser powers (L3 in red, L2 in black, L1 in blue and L0 in green). Cross markers denote experimental data and solid lines represent theoretical predictions. Plots are vertically shifted  for clarity. The parameters used for the calculation for the case of laser
on (off) are: $P_{z\mathrm{ST}}=-0.035$, $P_{z\mathrm{SO}}=-0.55$,
$\rho_{\mathrm{s}}=1.23\times10^{17}\operatorname{cm}^{-3}$, $T_{2}%
=219\operatorname{ns}$, $T_{1\mathrm{T}}=23\operatorname{ms}$ ($T_{1\mathrm{T}%
}=565\operatorname{ms}$) \cite{Loretz_064413} and $g_{\mathrm{s}}%
/2\pi=5.05\operatorname{Hz}$ ($g_{\mathrm{s}}/2\pi=2.72\operatorname{Hz}$).
The rate $T_{1\mathrm{O}}^{-1}$ of OISP is taken to be given by
$T_{1\mathrm{O}}^{-1}=0.16\times\gamma_{\mathrm{O}}$, where $\gamma
_{\mathrm{O}}=I_{\mathrm{L}}\sigma\lambda_{\mathrm{L}}/hc$ is rate of optical
absorption, where $\sigma=3\times10^{-17}\operatorname{cm}^{2}$
\cite{Wee_9379} is the optical cross section, $h$ is the Plank's constant and
$c$ is the speed of light in vacuum. The effective coupling coefficient
$g_{\mathrm{s}}$ for both cases of laser on and off is calculated using
Eq.~(\ref{g_s^2}) and the numerically calculated mode shape [see
Fig.\ref{Fig_resonator} (d)]. The volume inside the diamond wafer illuminated
by the laser is $0.76\operatorname{mm}^{3}$.}%
\label{Fig_ESR}%
\end{center}
\end{figure*}

The same two spin resonances seen in Fig.~\ref{Fig_diamond}(c) can be detected
without employing the technique of ODMR provided that their frequencies are
tuned close to the spiral resonator frequency $\omega_{\mathrm{c}}/2\pi$. The
plots (D: P1; L0), (D: P2; L0) and (D: P3; L0) of Fig.~\ref{Fig_ESR} depicts
NA measurements of the microwave reflectivity coefficient $R_{\mathrm{c}}$
with three different values of the injected signal microwave power
$P_{\mathrm{p}}$. No laser light is injected into the diamond wafer in these
measurements (labeled by L0 in Fig.~\ref{Fig_ESR}). Henceforth this method of
spin detection is referred to as cavity-based detection of magnetic resonance
(CDMR). Both CDMRs seen in Fig.~\ref{Fig_ESR} exhibit strong dependence on
$P_{\mathrm{p}}$, indicating thus that the interaction with the spins makes
the cavity response highly nonlinear.

To account for the observed spin-induced nonlinearity, the experimental
results are compared with theoretical predictions \cite{Boissonneault_060305}.
The decoupled cavity mode is characterized by an angular resonance frequency
$\omega_{\mathrm{c}}$, Kerr coefficient $K_{\mathrm{c}}$, linear damping rate
$\gamma_{\mathrm{c}}$ and cubic damping (two-photon absorption) rate
$G_{\mathrm{c}}$. The response of the decoupled cavity in the weak nonlinear
regime (in which, nonlinearity is taken into account to lowest non-vanishing
order) can be described by introducing the complex and mode amplitude
dependent cavity angular resonance frequency $\Upsilon_{\mathrm{c}}$, which is
given by%
\begin{equation}
\Upsilon_{\mathrm{c}}=\omega_{\mathrm{c}}-i\gamma_{\mathrm{c}}+\left(
K_{\mathrm{c}}-iG_{\mathrm{c}}\right)  E_{\mathrm{c}}\;, \label{Upsilon_c}%
\end{equation}
where $E_{\mathrm{c}}$ is the averaged number of photons occupying the cavity
mode. The imaginary part of $\Upsilon_{\mathrm{c}}$ represents the effect of
damping and the terms proportional to $E_{\mathrm{c}}$ represent the nonlinear
contribution to the response.

The effect of the spins on the cavity response in the weak nonlinear regime is
theoretically evaluated in \cite{Buks_033807}. The steady state cavity mode
response is found to be equivalent to the response of a mode having effective
complex cavity angular resonance frequency $\Upsilon_{\mathrm{eff}}$ given by
$\Upsilon_{\mathrm{eff}}=\Upsilon_{\mathrm{c}}+\Upsilon_{\mathrm{s}}$, where
$\Upsilon_{\mathrm{s}}=\sum_{n}\Upsilon_{n}$ and $\Upsilon_{n}$, which
represents the contribution of a spin labeled by the index $n$, is given by
[see Eq.~(4) in \cite{Buks_033807}]%
\begin{equation}
\Upsilon_{n}=-\frac{g_{n}^{2}}{\Delta_{n}}\frac{1-\frac{i}{\Delta_{n}T_{2,n}}%
}{1+\frac{1+4g_{n}^{2}T_{1,n}T_{2,n}E_{\mathrm{c}}}{\Delta_{n}^{2}T_{2,n}^{2}%
}}P_{z\mathrm{S},n}\;, \label{Upsilon_n}%
\end{equation}
where $g_{n}$ is the coupling coefficient between the $n$th spin and the
cavity mode, $T_{1,n}$ and $T_{2,n}$ are the spin's longitudinal and
transverse relaxation times, respectively, $\Delta_{n}=\omega_{\mathrm{c}%
}-\omega_{\mathrm{s},n}$ is the frequency detuning between the cavity
frequency $\omega_{\mathrm{c}}$ and the spin's transition frequency
$\omega_{\mathrm{s},n}$, and $P_{z\mathrm{S},n}$ is the spin's longitudinal
polarization. The term proportional to $E_{\mathrm{c}}$ in the denominator of
Eq.~(\ref{Upsilon_n}) gives rise to nonlinear response.

The coupling coefficients $g_{n}$ can be extracted from the numerically
calculated magnetic field induction $\mathbf{B}_{\mathrm{c}}\left(
\mathbf{r}\right)  $ of the spiral's fundamental mode [see
Fig.~\ref{Fig_resonator}(d)] using the expression $g_{n}=\gamma_{\mathrm{e}%
}\left\vert \mathbf{B}_{\mathrm{c}}\left(  \mathbf{r}_{n}\right)  \right\vert
\sin\varphi_{n}/E_{\mathrm{c}}^{1/2}$ \cite{Kubo_140502}, where $\mathbf{B}%
_{\mathrm{c}}\left(  \mathbf{r}_{n}\right)  $ is the cavity mode magnetic
induction at the location of the spin $\mathbf{r}_{n}$ and $\varphi_{n}$ is
the angle between $\mathbf{B}_{\mathrm{c}}\left(  \mathbf{r}_{n}\right)  $ and
the NV axis. When all contributing spins share the same detuning factor
$\Delta$, polarization$P_{z\mathrm{S}}$ and the same relaxation times $T_{1}$
and $T_{2}$, and when the variance in the distribution of $g_{n}$ is taken
into account to lowest nonvanishing order only, one finds that%
\begin{equation}
\Upsilon_{\mathrm{s}}=\frac{N_{\mathrm{eff}}g_{\mathrm{s}}^{2}}{\Delta}%
\frac{1-\frac{i}{\Delta T_{2}}}{1+\frac{1+\frac{E_{\mathrm{c}}}{E_{\mathrm{cc}%
}}}{\Delta^{2}T_{2}^{2}}}{}, \label{Upsilon_s}%
\end{equation}
where $\rho_{\mathrm{s}}\left(  \mathbf{r}\right)  $ is the density of
contributing NV$^{-}$ defects, $N_{\mathrm{eff}}=-\int d\mathbf{r}{}%
\rho_{\mathrm{s}}P_{z\mathrm{S}}$ is their effective number, the effective
coupling coefficient $g_{\mathrm{s}}$ is given by%
\begin{equation}
g_{\mathrm{s}}^{2}=\frac{\gamma_{\mathrm{e}}^{2}\mu_{0}\hbar\omega
_{\mathrm{c}}\int\mathrm{d}\mathbf{r}{}\rho_{\mathrm{s}}\left\vert
\mathbf{B}_{\mathrm{c}}\right\vert ^{2}\sin^{2}\varphi P_{z\mathrm{S}}}%
{\int\mathrm{d}\mathbf{r}{}\left\vert \mathbf{B}_{\mathrm{c}}\right\vert
^{2}\int\mathrm{d}\mathbf{r}{}\rho_{\mathrm{s}}P_{z\mathrm{S}}}{},
\label{g_s^2}%
\end{equation}
and $E_{\mathrm{cc}}=\left(  4g_{\mathrm{s}}^{2}T_{1}T_{2}\right)  ^{-1}$.

The underlying mechanism responsible for the spin-induced nonlinearity in the
cavity mode response is attributed to the change in spin polarization that
occurs via the cavity-mediated spin driving. As can be seen from Eq. (A83) of
Ref. \cite{Buks_033807}, the normalized change in polarization is proportional
to the ratio $E_{\mathrm{c}}/E_{\mathrm{cc}}$. Consequently, the induced
nonlinearity is expected to be negligibly small when $E_{\mathrm{c}}\ll
E_{\mathrm{cc}}$ [as is also seen from Eq. (\ref{Upsilon_s})]. On the other
hand, when $E_{\mathrm{c}}\gg E_{\mathrm{cc}}$ spin depolarization becomes
saturated. In this limit $\Upsilon_{\mathrm{eff}}=\Upsilon_{\mathrm{c}%
}+\Upsilon_{\mathrm{s}}\simeq\Upsilon_{\mathrm{c}}$ [see Eq. (\ref{Upsilon_s}%
)], and consequently the cavity mode is expected to become effectively
decoupled from the spins (this effective decoupling refers only to the
averaged response, whereas noise properties remain affected by the spins). The
regime of weak nonlinearity, in which nonlinearity can be taken into account
to lowest nonvanishing order only, is discussed in appendix A. Note,
however, that in the current experiment the nonlinearity can be considered as
weak only in a narrow region, and most observations cannot be properly
explained without accounting for higher order nonlinear terms.

%EndExpansion

In general, the averaged number of photons $E_{\mathrm{c}}$ is found from the
steady state solution of the equations of motion that govern the dynamics of
the system \cite{Buks_033807}. To lowest non-vanishing order in the coupling
coefficient $g_{\mathrm{s}}$ the effect of spins can be disregarded in the
calculation of $E_{\mathrm{c}}$. When, in addition, the intrinsic cavity mode
nonlinearity, which is characterized by the parameters $K_{\mathrm{c}}$ and
$G_{\mathrm{c}}$, has a negligibly small effect, the number $E_{\mathrm{c}}$
can be approximated by the following expression [see Eq.~(37) in
\cite{Yurke_5054}]%
\begin{equation}
E_{\mathrm{c}}=\frac{4\gamma_{\mathrm{f}}P_{\mathrm{p}}}{\hbar\omega
_{\mathrm{c}}}\frac{1}{\left(  \omega_{\mathrm{p}}-\omega_{\mathrm{c}}\right)
^{2}+\left(  \gamma_{\mathrm{f}}+\gamma_{\mathrm{c}}\right)  ^{2}}\;.
\label{E_c L}%
\end{equation}
As can be seen from Eq.~(\ref{Upsilon_s}), $\left\vert \Upsilon_{\mathrm{s}%
}\right\vert $ is a monotonically decreasing function of $E_{\mathrm{c}}$.
This suggests that the approximation in which Eq.~(\ref{E_c L}) is employed
for evaluating $E_{\mathrm{c}}$ (without taking into account both nonlinearity
and the coupling to the spins) remains valid even when $4g_{\mathrm{s}}%
^{2}T_{1}T_{2}E_{\mathrm{c}}\gg1$ provided that intrinsic cavity mode
nonlinearity remains sufficiently small. When intrinsic cavity mode
nonlinearity can be disregarded the cavity mode reflectivity $R_{\mathrm{c}}$
is given by \cite{Yurke_5054}%
\begin{equation}
R_{\mathrm{c}}=\frac{\left(  \omega_{\mathrm{p}}-\Omega_{\mathrm{c}}\right)
^{2}+\left(  \gamma_{\mathrm{f}}-\Gamma_{\mathrm{c}}\right)  ^{2}}{\left(
\omega_{\mathrm{p}}-\Omega_{\mathrm{c}}\right)  ^{2}+\left(  \gamma
_{\mathrm{f}}+\Gamma_{\mathrm{c}}\right)  ^{2}}\ , \label{R SNL}%
\end{equation}
where the real frequencies $\Omega_{\mathrm{c}}$ and $\Gamma_{\mathrm{c}}$ are
related to the complex frequency $\Upsilon_{\mathrm{eff}}$ by the relation
$\Upsilon_{\mathrm{eff}}=\Omega_{\mathrm{c}}-i\Gamma_{\mathrm{c}}$.

The fully-analytical theoretical predictions given by Eqs.~(\ref{Upsilon_s}),
(\ref{E_c L}) and (\ref{R SNL}) are employed for generating the plots (F: P1;
L0), (F: P2; L0) and (F: P3; L0) of Fig.~\ref{Fig_ESR}, which exhibit good
agreement with the corresponding CDMR data plots (D: P1; L0), (D: P2; L0) and
(D: P3; L0). The parameters that have been employed for the calculation are
listed in the figure caption. These findings support the hypothesis that the
above-discussed spin-induced nonlinearity is the underlying mechanism
responsible for the suppression of electron spin resonance (ESR) at relatively
high microwave input power $P_{\mathrm{p}}$.

The CDMR data plots in Fig.~\ref{Fig_ESR} labeled by L1, L2 and L3 are
obtained from measurements with laser intensities $2.15%
%TCIMACRO{\unit{mW}}%
%BeginExpansion
\operatorname{mW}%
%EndExpansion%
%TCIMACRO{\unit{mm}}%
%BeginExpansion
\operatorname{mm}%
%EndExpansion
^{-2}$, $12.8%
%TCIMACRO{\unit{mW}}%
%BeginExpansion
\operatorname{mW}%
%EndExpansion%
%TCIMACRO{\unit{mm}}%
%BeginExpansion
\operatorname{mm}%
%EndExpansion
^{-2}$ and $30%
%TCIMACRO{\unit{mW}}%
%BeginExpansion
\operatorname{mW}%
%EndExpansion%
%TCIMACRO{\unit{mm}}%
%BeginExpansion
\operatorname{mm}%
%EndExpansion
^{-2}$, respectively. As can be seen from the comparison to the plots labeled
by L0, in which the laser is turned off, the optical illumination strongly
affects the measured cavity response.

The laser-induced change in the cavity response is attributed to the mechanism
of OISP \cite{Loretz_064413, drake_013011,
Robledo_025013,Redman_3420,Harrison_586}. Spin is conserved in the optical
dipole transitions between the triplet ground state $^{3}A_{2}$ of NV$^{-}$
and the triplet first excited state $^{3}E$. However, transition from the spin
states $m_{\mathrm{e}}=\pm1$ of $^{3}E$ to the ground state is also possible
through an intermediate singlet states in a two-steps non-radiative process.
Such non-radiative process is also possible for the decay of the state
$m_{\mathrm{e}}=0$ of $^{3}E$, however, the probability of this process is
about $7$ times smaller than the probability of non-radiative decay of the
$m_{\mathrm{e}}=\pm1$ states \cite{Doherty_1}. The asymmetry between the decay
of $m_{\mathrm{e}}=0$ state, which is almost exclusively radiative, and the
decay of the states $m_{\mathrm{e}}=\pm1$, which can occur via non-radiative
process, gives rise to OISP. For our experimental conditions the probability
to find any given NV$^{-}$ defect at any given time not in the triplet ground
state $^{3}A_{2}$ is about $10^{-5}$ or less \cite{Doherty_1}. This fact is
exploited below for taking the effect of OISP into account within the
framework of a two-level model.

The effect of OISP can be accounted for by adjusting the values of the
longitudinal relaxation time $T_{1}$ and longitudinal steady state
polarization $P_{z\mathrm{S}}$ and make them dependent on laser intensity
$I_{\mathrm{L}}$. The total rate of spin longitudinal damping $\gamma
_{\mathrm{s}1}$ is given by \cite{Shin_124519}%
\begin{equation}
\gamma_{\mathrm{s}1}=-\frac{P_{z}-P_{z\mathrm{ST}}}{T_{1\mathrm{T}}}%
-\frac{P_{z}-P_{z\mathrm{SO}}}{T_{1\mathrm{O}}}\ , \label{gamma_a1 V1}%
\end{equation}
where the first term represents the contribution of thermal relaxation and the
second one represents the contribution of OISP. Here $P_{z}$ is the
instantaneous longitudinal polarization and $T_{1\mathrm{T}}^{-1}$
($T_{1\mathrm{O}}^{-1}$) is the rate of thermal relaxation (OISP). In steady
state and when $T_{1\mathrm{T}}^{-1}\gg T_{1\mathrm{O}}^{-1}$ (i.e. when OISP
is negligibly small) the coefficient $P_{z\mathrm{ST}}=-\tanh\left(
\hbar\omega_{\mathrm{s}}/2k_{\mathrm{B}}T\right)  $ is the value of $P_{z}$ in
thermal equilibrium, where $k_{\mathrm{B}}$ is the Boltzmann's constant and
where $T$ is the temperature. In the opposite limit of $T_{1\mathrm{O}}%
^{-1}\gg T_{1\mathrm{T}}^{-1}$ (i.e. when thermal relaxation is negligibly
small) the coefficient $P_{z\mathrm{SO}}$ is the value of $P_{z}$ in steady
state. Note that the total longitudinal damping rate $\gamma_{\mathrm{s}1}$
(\ref{gamma_a1 V1}) can be expressed as $\gamma_{\mathrm{s}1}=-T_{1}%
^{-1}\left(  P_{z}-P_{z\mathrm{S}}\right)  $, where $T_{1}^{-1}=T_{1\mathrm{T}%
}^{-1}+T_{1\mathrm{O}}^{-1}$ is the effective longitudinal relaxation rate,
and the effective steady state longitudinal polarization $P_{z\mathrm{S}}$ is
given by $T_{1}^{-1}P_{z\mathrm{S}}=T_{1\mathrm{T}}^{-1}P_{z\mathrm{ST}%
}+T_{1\mathrm{O}}^{-1}P_{z\mathrm{SO}}$.

The theoretical expressions given above for $T_{1}^{-1}$ and $P_{z\mathrm{S}}$
are employed for generating the plots labeled by F of Fig.~\ref{Fig_ESR} for
both cases of laser off (L0) and laser on (L1, L2 and L3). In spite of the
simplicity of the model that is employed for the description of OISP, good
agreement is obtained from the comparison with the CDMR data plots labeled by
D in a very wide range of values for the microwave power and laser intensity
(the entire explored range of $P_{\mathrm{p}}<0$ dBm and $I_{\mathrm{L}}<30%
%TCIMACRO{\unit{mW}}%
%BeginExpansion
\operatorname{mW}%
%EndExpansion%
%TCIMACRO{\unit{mm}}%
%BeginExpansion
\operatorname{mm}%
%EndExpansion
^{-2}$). Note that no resonance splitting is observed in all CDMR measurements.

The lineshapes of both ODMR and CDMR depend on the values of spin longitudinal $T_{1}$ and transverse
$T_{2}$ damping times. In order to check consistency we employ Eq. (2) of Ref. \cite{Jensen_014115} in order to express the full width at half minimum (FWHM)
$\Delta\nu$ of the ODMR in terms of $T_{1}$, $T_{2}$ and the driving
amplitude, which is denoted by $\omega_{1}$ ($\omega_{1}$ coincides with the
Rabi frequency at resonance). In the calculation of $\omega_{1}$ it is assumed
that the loop antenna can be treated as a perfect magnetic dipole. By
substituting the damping times $T_{1}$ and $T_{2}$ that are listed in the
caption of Fig.~\ref{Fig_ESR} into Eq. (2) of Ref. \cite{Jensen_014115} one
obtains $\Delta\nu=14%
%TCIMACRO{\unit{MHz}}%
%BeginExpansion
\operatorname{MHz}%
%EndExpansion
$, whereas the FWHM value extracted from the ODMR data using a fit to a
Lorentzian is $13.5%
%TCIMACRO{\unit{MHz}}%
%BeginExpansion
\operatorname{MHz}%
%EndExpansion
$.

\begin{figure}
[ptb]
\begin{center}
\includegraphics[
height=2.0003in,
width=3.4in
]%
{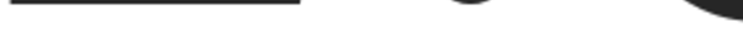}%
\caption{Cavity mode reflectivity $R_{\mathrm{c}}$ with P1 defects for various
values of injected microwave power $P_{\mathrm{p}}$ ($\mathrm{P}%
1=-90~\mathrm{dBm}$, $\mathrm{P}2=-80~\mathrm{dBm}$ and $\mathrm{P}%
3=-70~\mathrm{dBm}$) with laser off. For each pair the top plot is
experimental data (labeled by D) and the bottom is the theoretical prediction
(labeled by F). The parameters used for the calculation are: $\gamma
_{\mathrm{c}}/2\pi=0.304\operatorname{MHz}$, $\gamma_{\mathrm{f}}%
/2\pi=0.349\operatorname{MHz}$, $\rho_{\mathrm{s}}=1\times10^{18}%
\operatorname{cm}^{-3}$, $T_{2}=438\operatorname{ns}$ and $T_{1\mathrm{T}%
}=470\operatorname{ms}$ \cite{reynhardt1998}. The values of $P_{z\mathrm{ST}}$
and $g_{\mathrm{s}}$ are the same as in Fig. \ref{Fig_ESR} for the case of
laser off.}%
\label{Fig_P1}%
\end{center}
\end{figure}

A CQED resonance due to P1 defects \cite{kaiser1959,smith1959} is observed
when the externally applied magnetic field is tuned close to the value of $89%
%TCIMACRO{\unit{mT}}%
%BeginExpansion
\operatorname{mT}%
%EndExpansion
$\ (see Fig.~\ref{Fig_P1}). When both nuclear Zeeman shift and nuclear
quadrupole coupling are disregarded, the spin Hamiltonian of a P1 defect is
given by \cite{Loubser_1201,Schuster_140501,Wood_155402} $\mathcal{H}%
=\gamma_{\mathrm{e}}\mathbf{B}\cdot\mathbf{S}+\hbar^{-1}A_{\perp}\left(
S_{x}I_{x}+S_{y}I_{y}\right)  +\hbar^{-1}A_{\parallel}S_{z}I_{z}$, where
$\mathbf{S}=\left(  S_{x},S_{y},S_{z}\right)  $ is an electronic spin 1/2
vector operator, $\mathbf{I}=\left(  I_{x},I_{y},I_{z}\right)  $ is a nuclear
spin 1 vector operator, $A_{\parallel}=2\pi\times114.03%
%TCIMACRO{\unit{MHz}}%
%BeginExpansion
\operatorname{MHz}%
%EndExpansion
$ and $A_{\perp}=2\pi\times81.33%
%TCIMACRO{\unit{MHz}}%
%BeginExpansion
\operatorname{MHz}%
%EndExpansion
$ are respectively the longitudinal and transverse hyperfine parameters, and
the $z$ direction corresponds to the diamond $\left\langle 111\right\rangle $
axis. When the externally applied magnetic field $\mathbf{B}$ is pointing
close to a crystal direction $\left\langle 100\right\rangle $, i.e. when
$\cos^{2}\theta\simeq1/3$, the electron spin resonance at angular frequency
$\gamma_{\mathrm{e}}B$ is split due to the interaction with the nuclear spin
into three resonances, corresponding to three transitions, in which the
nuclear spin magnetic quantum number is conserved. To first order in
perturbation theory the angular resonance frequencies are given by
$\gamma_{\mathrm{e}}B-\omega_{\mathrm{en}}$, $\gamma_{\mathrm{e}}B$ and
$\gamma_{\mathrm{e}}B+\omega_{\mathrm{en}}$, where $\omega_{\mathrm{en}}%
^{2}=A_{\parallel}^{2}\cos^{2}\theta+A_{\perp}^{2}\sin^{2}\theta$. For the
case where $\cos^{2}\theta=1/3$ the calculated splitting is given by
$\omega_{\mathrm{en}}/2\pi=93.5%
%TCIMACRO{\unit{MHz}}%
%BeginExpansion
\operatorname{MHz}%
%EndExpansion
$, whereas the value extracted from the data seen in Fig.~\ref{Fig_P1} is
$93.82%
%TCIMACRO{\unit{MHz}}%
%BeginExpansion
\operatorname{MHz}%
%EndExpansion
$. The plots labeled by F in Fig. \ref{Fig_P1} represent the theoretical
prediction based on the analytical expressions (\ref{Upsilon_s}),
(\ref{E_c L}) and (\ref{R SNL}). The comparison with the CDMR data plots
(labeled by D) yields a good agreement. The parameters that have been employed
for the calculation are listed in the figure caption.%

%TCIMACRO{\FRAME{ftbpFU}{3in}{2.0003in}{0pt}{\Qcb{Cavity mode reflectivity
%$R_{\mathrm{c}}$ with P1 defects for various values of injected microwave
%power $P_{\mathrm{p}}$ ($\mathrm{P}1=-90~\mathrm{dBm}$, $\mathrm{P}%
%2=-80~\mathrm{dBm}$ and $\mathrm{P}3=-70~\mathrm{dBm}$) with laser off. For
%each pair the top plot is experimental data (labeled by D) and the bottom is
%the theoretical prediction (labeled by F). The parameters used for the
%calculation are: $\gamma_{\mathrm{c}}/2\pi=0.304\operatorname{MHz}$,
%$\gamma_{\mathrm{f}}/2\pi=0.349\operatorname{MHz}$, $\rho_{\mathrm{s}}%
%=1\times10^{18}\operatorname{cm}^{-3}$, $T_{2}=438\operatorname{ns}$ and
%$T_{1\mathrm{T}}=470\operatorname{ms}$ \cite{reynhardt1998}. The values of
%$P_{z\mathrm{ST}}$ and $g_{\mathrm{s}}$ are the same as in Fig. \ref{Fig_ESR}
%for the case of laser off.}}{\Qlb{Fig_P1}}{figp1.eps}%
%{\special{ language "Scientific Word";  type "GRAPHIC";
%maintain-aspect-ratio TRUE;  display "ICON";  valid_file "F";  width 3in;
%height 2.0003in;  depth 0pt;  original-width 0pt;  original-height 0pt;
%cropleft "0";  croptop "1";  cropright "1";  cropbottom "0";
%filename 'FigP1.eps';file-properties "XNPEU";}} }%
%BeginExpansion
%EndExpansion

The nonlinearity in cavity response has an important impact on sensitivity of
spin detection. Let $S_{\mathrm{N}}$ be the minimum detectable change in the
number of spins $\delta N_{\mathrm{s}}$ per a given square root of the
available bandwidth (i.e. the inverse of the averaging time). It is assumed
that sensitivity is limited by the fundamental bound imposed upon the signal
to noise ratio by shot noise. When the cavity's response is linear
$S_{\mathrm{N}}$ is proportional to $E_{\mathrm{c}}^{-1/2}$ [see Eq.~(1) in
\cite{Buks_026217}], and thus in this regime sensitivity can be enhanced by
increasing the energy stored in the cavity $E_{\mathrm{c}}\hbar\omega
_{\mathrm{c}}$. However, nonlinearity, which can be avoided only when
$E_{\mathrm{c}}\ll E_{\mathrm{cc}}$ [see Eq.~(\ref{Upsilon_s})], imposes a
bound upon sensitivity enhancement. When the sensitivity coefficient
$S_{\mathrm{N}}$ is calculated according to Eq.~(1) in Ref. \cite{Buks_026217}
for the case where the number of cavity photons is taken to be $E_{\mathrm{cc}%
}$ and the responsivity is calculated using Eq. (\ref{omega_cs}) below, one
finds that $S_{\mathrm{N}}$ becomes%
\begin{equation}
S_{\mathrm{N}}\simeq\frac{2}{\left\vert P_{z\mathrm{ST}}\right\vert ^{3/2}%
}\left(  \frac{\gamma_{\mathrm{c}}}{g_{\mathrm{s}}^{2}}\frac{2T_{1}}{T_{2}%
}\right)  ^{1/2}\ . \label{S_N LL}%
\end{equation}
Note that in general $2T_{1}/T_{2}\geq1$ [see Eq.~(A79) in \cite{Buks_033807}%
]. For example, for the parameters of our device with laser off
Eq.~(\ref{S_N LL}) yields $S_{\mathrm{N}}=5\times10^{7}%
%TCIMACRO{\unit{Hz}}%
%BeginExpansion
\operatorname{Hz}%
%EndExpansion
^{-1/2}$. The estimate given by Eq.~(\ref{S_N LL}) is expected to be
applicable for any cavity-based technique of spin detection.

%When $E_{\mathrm{c}}\gtrsim E_{\mathrm{cc}}$ nonlinearity cannot be avoided.
%However, in some cases nonlinearity can be exploited for enhancing
%sensitivity. For example, nonlinearity in microwave resonators can be
%exploited for signal amplification \cite{Yurke_5054}. Large gain can be
%obtained when the resonator is externally driven close to the onset of
%bistability. In Ref. \cite{Yurke_5054} the performance of such an amplifier
%has been theoretically investigated based on the assumption that nonlinearity
%can be taken into account to lowest non-vanishing order only. However, this
%approximation becomes invalid when near the onset of bistability higher order
%terms become important. At the onset of bistability the averaged number
%$E_{\mathrm{c}}$ obtains a value $E_{\mathrm{co}}$ given by Eq.~(42) in Ref.
%\cite{Yurke_5054}. For the parameters of our device we find that
%$E_{\mathrm{co}}\simeq2E_{\mathrm{cc}}$. This estimate suggests that the
%spin-induced nonlinearity in our device is too strong and it is not suitable
%for the implementation of the signal amplification scheme proposed in
%\cite{Yurke_5054}, since the condition $E_{\mathrm{c}}\ll E_{\mathrm{cc}}$
%cannot be satisfied in the region of bistability.

To conclude, in this work we have observed strong coupling (i.e. cooperativity
larger than unity) between a superconducting microwave cavity and spin
ensembles in diamond (the measured values of the cooperativity parameter
$N_{\mathrm{eff}}g_{\mathrm{s}}^{2}/\gamma_{\mathrm{c}}\gamma_{2}$ are $14$
with the NV$^{-}$ ensemble and laser intensity of $30%
%TCIMACRO{\unit{mW}}%
%BeginExpansion
\operatorname{mW}%
%EndExpansion%
%TCIMACRO{\unit{mm}}%
%BeginExpansion
\operatorname{mm}%
%EndExpansion
^{-2}$ and $6.2$ with the P1 ensemble). We find that the coupling imposes an
upper bound upon the input microwave power, for which the cavity response
remains linear. This bound has important implications on the sensitivity of
traditional spin detection protocols that are based on linear response. On the
other hand, in some cases nonlinearity can be exploited for sensitivity
enhancement (e.g. by generating parametric amplification). However, further
study is needed to explore ways of optimizing the performance of sensors
operating in the nonlinear regime.

We thank Adrian Lupascu for useful discussions. This work is supported by the
Israeli Science Foundation and the Binational Science Foundation.

\appendix

\section{Weak Nonlinearity}

In the weak nonlinear regime it is assumed that the averaged number of cavity
mode photons $E_{\mathrm{c}}$ is sufficiently small to allow taking
nonlinearity into account to lowest non-vanishing order only. In this limit
the cavity mode has a nonlinear response that can be adequately described
using the well-known Duffing/Kerr model \cite{Roy_740}. However, as is
discussed below, when higher order terms in $E_{\mathrm{c}}$ become
significant the response can no longer be described by the Duffing/Kerr model.
The distinction becomes most pronounced in the limit of high input microwave
power. Both our experimental (see Figs. \ref{Fig_ESR} and \ref{Fig_P1}) and
theoretical [see Eq. (\ref{Upsilon_s})] results indicate that the cavity mode
becomes effectively decoupled from the spins in the limit of high microwave
power. Consequently linearity is restored at high input power, provided that
the input power is not made too high and it is kept below the region where
intrinsic cavity mode nonlinearity, which is characterized by the intrinsic
Kerr coefficient $K_{\mathrm{c}}$ and intrinsic cubic damping rate
$G_{\mathrm{c}}$, becomes significant. Note that in our device the intrinsic
nonlinearity becomes noticeable only when the input power exceeds a value of
about $0$ dBm, which is 5-6 orders of magnitude higher than the value at which
the cavity becomes effectively decoupled from the spins.

To first order in $E_{\mathrm{c}}$ the spin-induced shift $\Upsilon
_{\mathrm{s}}$ in the complex cavity mode angular frequency can be expanded as
[see Eq. (\ref{Upsilon_s})]%
\begin{equation}
\Upsilon_{\mathrm{s}}=\omega_{\mathrm{cs}}-i\gamma_{\mathrm{cs}}+\left(
K_{\mathrm{cs}}-iG_{\mathrm{cs}}\right)  E_{\mathrm{c}}+O\left(
E_{\mathrm{c}}^{2}\right)  {}, \label{Upsilon_s 1st order}%
\end{equation}
where the shift in linear frequency $\omega_{\mathrm{cs}}$ and the Kerr
coefficient $K_{\mathrm{cs}}$ are given by%
\begin{align}
\omega_{\mathrm{cs}}  &  =\frac{N_{\mathrm{eff}}g_{\mathrm{s}}^{2}}{\Delta
}\frac{1}{1+\zeta_{2}^{2}}{},\label{omega_cs}\\
K_{\mathrm{cs}}  &  =-\frac{N_{\mathrm{eff}}g_{\mathrm{s}}^{2}}{\Delta
E_{\mathrm{cc}}}\left(  \frac{\zeta_{2}}{1+\zeta_{2}^{2}}\right)  ^{2}{},
\label{K_cs}%
\end{align}
the linear damping rate is given by $\gamma_{\mathrm{cs}}=\zeta_{2}%
\omega_{\mathrm{cs}}$ and the cubic damping rate is given by $G_{\mathrm{cs}%
}=\zeta_{2}K_{\mathrm{cs}}$, and where $\zeta_{2}=1/\Delta T_{2}$. In the
regime of linear response (i.e. when $\Upsilon_{\mathrm{s}}=\omega
_{\mathrm{cs}}-i\gamma_{\mathrm{cs}}$) Eq. (\ref{Upsilon_s 1st order})
reproduces well-known results for spin-induced frequency shift and
broadening of the cavity resonance \cite{Haroche_24}. The validity conditions
for Eqs. (\ref{Upsilon_s}) and (\ref{Upsilon_s 1st order}) are discussed in
Ref. \cite{Buks_033807}.

In general, in the weak nonlinear regime, in which higher order terms in
$E_{\mathrm{c}}$ can be disregarded, the terms proportional to $E_{\mathrm{c}%
}$ in the complex angular frequency shift $\Upsilon_{\mathrm{s}}$
[see Eq. (\ref{Upsilon_s 1st order})] may give rise to bistability in the response of
the system to an applied monochromatic driving. At the onset of bistability
the averaged number $E_{\mathrm{c}}$ obtains a value denoted by
$E_{\mathrm{co}}$. When the value of $E_{\mathrm{co}}$ is estimated based on
the assumption that higher order terms in $E_{\mathrm{c}}$ may be disregarded
one finds for the parameters of our device that $E_{\mathrm{co}}\simeq
2E_{\mathrm{cc}}$ (calculated using Eq. (42) in Ref. \cite{Yurke_5054}). On
the other hand, the assumption that higher order terms in $E_{\mathrm{c}}$ may
be disregarded is applicable only when $E_{\mathrm{c}}\ll E_{\mathrm{cc}}$, and thus the nonlinearity cannot be considered as weak in this region.
When the bistability is accessible the system can be used for signal amplification \cite{Roy_740}, which can yield a significant gain close to the onset of bistability \cite{Yurke_5054}.

%\bibliography{Eyal_Bib_C}
%\bibliographystyle{apsrev}
%\bibliography{Eyal_Bib}

\end{document}